\documentclass[a4paper,12pt]{article}
\usepackage{amsmath,amssymb}
\usepackage{graphicx}   

\title{Three-dimensional localized coherent structures of surface
turbulence. III~Experiment and model validation.}
\author{E.~A. Demekhin $^{[1]}$, E.~N. Kalaidin and
A.~S. Selin \\
Kuban State University\\
Stavropolskaya st., 149\\
Krasnodar 350040, Russia}

\begin{document}

\setlength{\baselineskip}{18pt} \maketitle
\footnotetext[1]{E-mail: edemekhi@gmail.com}

\begin{center}
Abstract
\end{center}

The paper continues a series of publications devoted to the 3D
nonlinear localized coherent structures on the surface of
vertically falling liquid films. The work is primarily focussed on
experimental investigations. We study: (i) instabilities and
transitions leading to 3D  coherent structures; (ii)
characteristics of these  structures. Some nonstationary effects
are also studied numerically. Our experimental results, as well as the
results of other investigators, are in a good agreement with our
theoretical and numerical predictions.

\newpage
\setlength{\baselineskip}{24pt}
\section{Introduction}
\renewcommand{\thesubsection}{\Alph{subsection}}
\subsection{Motivation}

Localized nonlinear coherent structures play an important role in
many phenomena connected with the chaotic motion of liquid, in
particular, in some regimes of falling films. As we discussed in
Part I $^{\cite{D1}}$, the analysis of experiments reveals the
existence of several wave regimes, depending on the inlet Reynolds
number $\langle Re \rangle$, $(0,\langle Re^{(1)}\rangle)$,
$(\langle Re^{(1)}\rangle,\langle Re^{(2)}\rangle)$ and  $(\langle
Re^{(2)}\rangle,\langle Re^{(3)}\rangle)$. For water, $\langle
Re^{(1)}\rangle \approx 3-5$, $\langle Re^{(2)}\rangle \approx
40-70$ and $\langle Re^{(3)}\rangle \approx 400$.  In the first
interval the instability is too weak for any manifestation to be
seen and the surface is flat, in the second interval the waves are
two-dimensional (2D) and in the third interval the surface is
covered with three-dimensional (3D) localized coherent structures,
$\Lambda$-solitons. These 3D solitons are stable and robust and
are involved in continuous chaotic movement, interacting with
each other as ``quasi-particles''. This is the so-called regime of
``surface turbulence'' (ST). Note that describing ST in terms of the
motion of ``quasi-particles'' results in a dramatic
reduction in the number of degrees of freedom.

In our theoretical investigation $^{\cite{D1,D2}}$ we
systematically apply the simplified system of equations with
respect to local film thickness and flow rates in the direction
of the flow and in the transverse direction, the Kapitsa-Shkadov
system. For large Kapitsa number $\gamma=\sigma \rho^{-1}
\nu^{-4/3} g^{-1/3}$ (where $\sigma$ is surface tension, $\rho$
and $\nu$ are the liquid density and viscosity  and $g$ is the
acceleration of gravity) the flow is described by the universal
parameter $\delta$ (the modified Reynolds number) and the results
can be re-calculated for any liquid using the relation
$Re=7.51\gamma^{3/11} \delta^{9/11}$. In particular, for water,
$Re= 66.1 \delta^{9/11}$. As $Re \to 0$, the Kapitsa-Shkadov system
collapses into the Kuramoto-Sivashinsky (KS) equation.

In Part~I $^{\cite{D1}}$,  we have examined theoretically the
instability of 2D waves with respect to a disturbances in the
transverse direction. We have also established the 2D--3D
transition scenarios resulting in fully developed 3D waves whose
complex interaction leads eventually to ST. In Part~II
$^{\cite{D2}}$, we considered 3D coherent structures of the ST
as stationary running 3D solitary pulses, referred to as
$\Lambda$\nobreakdash-solitons. We have found these nonlinear
solutions  and have  studied their linear stability. It has been
found that for the universal parameter in the range $\delta<0.054$
the $\Lambda$-solitons are unstable
with respect to the continuous part of spectrum ("radiation
instability") and, at $\delta>0.51$, $\Lambda$-solitons do not
exist. Hence their region of existence/stability is
$0.054<\delta<0.51$ which for water translates to $6<Re<38$.

Among numerous works devoted to the experimental investigation of wave
dynamics in falling films, only works by Park and Nosoko
$^{\cite{P1}}$ and Alekseenko et al. $^{\cite{A7,A8}}$ are devoted
to  transitions to the regime of $\Lambda$-solitons and to the properties
of these structures. Note that for films on an inclined plane
with a small inclination angle, 3D solitary waves have not been observed
(see papers by Liu et al.$^{\cite{LSG}}$ ).

In the work $^{\cite{P1}}$ water was the working liquid. The
authors used artificial perturbations in the transverse direction
to destroy 2D solitary waves, the critical Reynolds number for the
transition to 3D waves was evaluated as  $\langle Re \rangle
\approx 40$. At this Reynolds number, the most dangerous length of
transverse perturbations was about 2 cm. In this work, it was
stated that disintegration of 2D solitons leads to 3D
solitary waves. The work contains many nice photographs of the
wavy surface, but, unfortunately, quantitative comparison with
theory is rather difficult: i) a strong influence from the wave
interaction complicates the analysis of the data, ii) the authors were
using the inlet Reynolds number $\langle Re \rangle$, while for
the comparison we need the substrate Reynolds number $Re$.
However, there is a qualitative correspondence between the
experimental data and our theory.

As a matter of fact, there are serious difficulties with
trying to single out an individual ``particle'' of ST and its
parameters from experiments. The main such difficulties are:
a)~the ``particles'' are the result of a long and complex
evolution downstream; b)~in the ST regime, the ``particles'' are
continuously interacting with each other and it is very difficult
if not impossible to measure the parameters of an individual
particle; sometimes it is difficult even to identify them in a
randomly disturbed surface.

In the experiments of Alekseenko et al. $^{\cite{A7,A8}}$, the
evolution of 3D localized signals was considered for substrate
Reynolds numbers $Re=1.25-4.7$ in an alcohol-water mixture with
$\gamma=404$ so that the interval of the universal modified
Reynolds number is $\delta=0.015-0.076$. In these studies, the
authors excited the 3D waves by a short-duration impact with a thin
jet of working fluid in the upper part of the film flow. The
major trends of 3D localized signals were presented and the existence of
stationary running 3D solitary waves was proved. For the time
being, only these experiments contain enough information to be
compared with the theoretical prediction.

The main aim of the present work is an experimental investigation
of some 3D phenomena in vertically falling films and a validation of
the model developed in Parts I and II. The range of Reynolds
numbers $\langle Re \rangle$ we wish to investigate is 5 to 100
(smaller Reynolds numbers were investigated by Alekseenko et
al.$^{\cite{A7,A8}}$). First, we study the 2D--3D transitions
using method $^{\cite{P1}}$ and perturbing 2D solitary waves in the
transverse direction. Second, we artificially create 3D solitary
waves by depositing a single drop on the undisturbed liquid surface.
A critical aspect of the experiment is that, in contrast to the works
$^{\cite{A7,A8}}$, the mass of the deposited drop corresponds to
the mass of 3D solitary wave obtained theoretically in Part II. To
complete the picture of evolution, some nonstationary effects are
studied numerically. The results of our experiments along with the
experiments of Alekseenko et al. $^{\cite{A7,A8}}$ are compared
with our theory in Parts I and~II.

\subsection{Experimental methodology}
  We are going to investigate the 2D-3D transition and $\Lambda$-solitons
for inlet Reynolds numbers $\langle Re \rangle$ from 5 to 100. For this
purpose,  we choose a planar vertical channel which is 25~cm long
and 15~cm wide and made of a special mirrored glass. To measure the
wave characteristics, we choose the fluorescent imaging method by
Liu and Gollub  $^{\cite{L4}}$. The method was also successfully
used by Vlachogiannis and Bontozoglou $^{\cite{V1}}$.

In order to investigate the transition from the regime of 2D solitary
waves to $\Lambda$-solitons (2D-3D transition), we applied the ideas of
$^{\cite{A5,L2}}$ and $^{\cite{P1}}$: regular 2D solitary waves
are created by low-frequency flow rate pulsations; these waves are
perturbed in the transverse direction by a system of special
needles separated by a certain distance. The
spatio-temporal evolution of such waves are considered in order to
find conditions of 3D instability and transition to 3D solitary
waves.

We  devise a novel direct method that allows us to create
individual $\Lambda$\nobreakdash-structures in our experiments.
The basic idea  is to deposit drops having the soliton's mass onto
the liquid substrate. The mass of the solitary wave is taken from
our theoretical calculations $^{\cite{D2}}$. After the drop
touches the surface, it is subsequently involved in film
movement and is rapidly transformed into a stationary running
solitary wave.

\subsection{Outline}
In Sec.~2, we obtain the connection between the inlet or pump
Reynolds number, $\langle Re \rangle$, frequently used in
experiments, and the Reynolds number based on the soliton's
substrate, $Re$, used in our theoretical investigation
$^{\cite{D1}}$. The ratio $\langle Re \rangle / Re$ depends on the
properties of the (3D or 2D) solitary wave, namely its volume and
velocity, as well as on the ``density'' of solitons or,
equivalently, the number of solitons per unit area. In the regime
of ST, this ratio is about 10.

In Sec.~3, we describe our simple experimental setup and a
method to measure wave characteristics. The planar vertical
channel is made of a special mirrored glass and distilled water is
used as working liquid. Experimental methods to investigate
transitions leading to 3D solitary waves and to create regular
3D waves on the film surface are discussed. The
numerical method used  to study the evolution of localized signals is
formulated.

In Sec.~4, we present our experimental results and compare them
with the theoretical ones. First, some visual observations of
natural waves and of their transitions are presented. Second,
normal 2D solitary waves, created by periodic pulsations of flow rate,
are considered. They are excited by special needles
in the transverse direction. The evolution downstream of these
perturbed 2D solitary waves can result in their disintegration and
further formation of 3D solitary waves. The most dangerous wave length
is found. Third, in order to create an individual 3D soliton, a
drop of liquid with a mass corresponding to the equilibrium
soliton is deposited directly onto the liquid substrate. Downstream,
the drop spreads out and takes the form of a stationary
running $\Lambda$-soliton. We also perform computations to model
the creation of the $\Lambda$ solitary wave from a localized
initial signal, or ``drop''. Good agreement between the theory in
Parts I and II and our experiments along with Alekseenko et al.
$^{\cite{A7,A8}}$ is demonstrated. Finally, we consider a
numerical solution of simple wave interactions.

Sec. 5 contains the main conclusions of the work and a
discussion of unsolved problems for future investigation.

\section{Connection between inlet and substrate Reynolds numbers}
The natural control parameter in experiments is the pump
discharge flow rate. The associated dimensionless parameter is the inlet
Reynolds number, $\langle Re \rangle$, which is the pump discharge
per unit channel width divided by the kinematic viscosity of the
liquid. This Reynolds number is used in the majority of
experimental studies of falling films. For fixed liquids, it is
the only free parameter.

On the other hand, when we consider individual localized
structures, it is convenient to introduce another Reynolds number,
the Reynolds number of the soliton substrate, $Re$: as we first
pointed out in Sec. II of Part I and Sec. III of Part I, far from
its hump a solitary wave decays rapidly to a flat flow, the so-called
``substrate'' or ``sublayer''. $Re$ is based on the substrate's
thickness. Note that the substrate Reynolds number has been used
in the past by several authors, see for example $^{\cite{TSV}}$.

 Here, we are interested in the
connection between the two Reynolds numbers, $\langle Re \rangle$
and $Re$. In the case of wave-less flow, the two numbers are the
same, $\langle Re \rangle$=$Re$, but for a wavy regime the
situation is quite different. In this case, the connection between
the two Reynolds numbers depends on the number of 2D or 3D
solitary waves per unit area in the $(x,z)$-plane or, equivalently,
the ``density'' of these ``quasi-particles''.

Let us first define the volumes of 2D and 3D solitary waves,
respectively,
\begin{equation}
 J_{2D}=\int_{-\infty}^{\infty} (h-1)dx
\end{equation}
and
\begin{equation}
 J_{3D}=\int_{-\infty}^{\infty} \int_{-\infty}^{\infty}(h-1)dx dz,
\end{equation}
where (2) was first introduced in Eq. (19) of Part II but is
rewritten here for clarity. As we pointed out in Sec. II D of Part
II, even though at large $\delta$ the amplitude of 3D solitons
tends to a constant, their volume $J_{3D}$ continues to increase.
For 2D solitons, we have similar behaviour, i.e. for large $\delta$
their amplitude saturates but their volume continues to increase
with $\delta$ due to the widening of their humps. In fact, for
large $\delta$, the volume $J_{2D}$ of 2D solitons is a linear
function of $\delta$, much like the volume of 3D solitons, and can
be well approximated by the relationship
\[
J_{2D}=208\delta+0.535
\]
For completeness, we also give the volume of 3D solitons for large
$\delta$ from Sec. II D in Part II:
\[
J_{3D}=2034\delta-81.54
\]

Let us assume that we have a wave-less substrate of unitary
thickness with Reynolds number $Re$ (the corresponding dimensional
thickness is $\tilde{h_0}$ --- see Sec. II of Part I). Let us also
assume that a group of 3D localized coherent structures,
$\Lambda$\nobreakdash-waves, with average distances between them
$L_x$ and $L_z$, in the streamwise and spanwise directions
respectively, slides with a speed $c$ along this substrate. We
also define the wave numbers, $\alpha=2\pi/L_x$ and
$\beta=2\pi/L_z$ in the $x$- and $z$-directions respectively. Let us
single out one ``quasi-particle'' and integrate the mass balance
equation (5c) in Part I in the frame moving with speed $c$ along a
period in the $z$-direction and use the fact that the $z$-component of
the flow rate, $p$, is zero far from the hump,
\begin{equation}
\frac{\partial}{\partial x}(Q-
 \frac{c}{L_z}\int_{-L_z/2}^{+L_z/2}h dz)=0
\end{equation}
where $Q=Q(x)$ is a averaged with respect to the $z$ flow rate,
\[
Q=\frac{1}{L_z}\int_{-L_z/2}^{+L_z/2}q dz
\]
Integrating (3) once with respect to $x$ gives
\begin{equation}
Q- \frac{c}{L_z}\int_{-L_z/2}^{+L_z/2}(h-1) dz=1
\end{equation}
where the integration constant was determined from the condition
that the soliton decays rapidly far from its hump to
plane-parallel flow $h=1, q=1$ (Eq. (7), Part~I). Further, we
integrate (4) once with respect to $x$:

\begin{equation}
\langle Q \rangle =1+\frac{c}{L_x L_z}\int_{-L_x/2}^{+L_x/2}
\int_{-L_z/2}^{+L_z/2}(h-1) dx dz
\end{equation}
where
\[
\langle Q \rangle=\frac{1}{L_x} \int_{-L_x/2}^{+L_x/2} Q dx
\]
is an averaged flow rate along the $x$-direction. If $L_x$ and
$L_z$ are sufficiently large, the double integral on the
righthand side of (5) can be approximated by $J_{3D}$ in (2),
giving the final relationship
\begin{equation}
\langle Q \rangle \sim 1+\frac{c J_{3D}}{L_x L_z}=1+\frac{\alpha
\beta c J_{3D}}{4\pi^2}.
\end{equation}

 The first term is the contribution
from the slowly moving thin substrate; the second term is the
contribution from the rapidly moving large hump. It depends on the
``density'' of pulses $\alpha \beta$, velocity of the wave $c$ and
its volume $J_{3D}$. Mass conservation between the flat inlet
region and the region of well-developed waves further downstream
implies that $\langle Q \rangle$ is equal to the dimensionless inlet
flow rate. Hence, the inlet Reynolds number $\langle Re \rangle$,
based on inlet flow rate, and the sublayer $Re$ based on
sublayer flow rate (whose dimensionless value $=1$), are
connected by the relation
\begin{equation}
\langle Re\rangle=\langle Q \rangle Re.
\end{equation}
We can also derive for the 2D wave regime an expression similar to
(5):
\begin{equation}
\langle Q \rangle =1+\frac{c J_{2D}}{L_x}=1+\frac{\alpha c
J_{2D}}{2\pi}
\end{equation}

From Eqs. (5) and (6), in the 3D-soliton regime, if either volume
$J_{3D} \to 0$ or ``density'' $\alpha \beta \to 0$ (equivalently
``volume'' $J_{2D} \to 0$ or ``density'' $\alpha  \to 0$ in the
2D-soliton regime), then $\langle Q \rangle \to 1$ and $\langle
Re\rangle \to Re$, as would be the case  for a single soliton
in an infinite domain. However, usually the contribution of the
second tern in (5) is much larger than the first term and  $\langle
Re \rangle$ is much larger than $Re$.

 The boundary of the 2D--3D transition in experiments  is about $\langle Re
\rangle \approx 40-70$ for water (see Fig. 1, Part I). We should
compare these values with our theoretical critical $\delta_* =
0.048$ or $Re_*=5.5$ for water (Sec. IV, Part I), which correspond
to instability of 2D solitons. In order to do this, we should
recalculate $Re$ as $\langle Re \rangle$ and apply relations
(5)--(7). However, this can only be done qualitatively because the
``density'' of the soliton gas and hence $\langle Q \rangle$ is
unknown. Nevertheless, we can use the results of previous experiments to estimate
the second term in (5) and hence evaluate $\langle Q \rangle$.
According to Chu and Dukler $^{\cite {C10}}$, the contribution of
the second term can be ten to twenty times larger than that of the
first one, so the majority of the inlet flow rate is carried
by solitons. For our range of Reynolds numbers, we take the second
term to be ten times the first one (see Fig. 12 of Chu and Dukler's
paper). This gives us the rough estimate
\begin{equation}
\langle Re \rangle=10Re.
\end{equation}
Hence, our transition Reynolds number $Re=5.5$ is equivalent to
$\langle Re \rangle=55$, which is in good agreement with the
experimental value $\langle Re^{(2)} \rangle \approx 40$--70 where the
2D--3D transition takes place (Fig.1, Part I).

An important related question is which type of solitons, 3D or 2D,
carries more liquid. This question, in particular, has important
consequences
 for heat/mass transfer. As we demonstrated in Figs. 11 and 12 of
 Part II, both the amplitude and speed of a 3D soliton are smaller than
 those of a 2D soliton of the same $\delta$.
On the other hand, as $\delta$ increases, a 3D soliton becomes
bulkier, and 3D solitons can potentially carry more liquid than 2D
solitions for the same $\delta$. By equating (8) to (6), we obtain the
limiting value $\beta_*$,
\begin{equation}
\beta_*=2\pi\frac{\alpha_{2D}}{\alpha_{3D}} \frac{c_{2D}}{c_{3D}}
\frac{J_{2D}}{J_{3D}}.
\end{equation}
If $\beta > \beta_*$, 3D solitons carry more liquid than 2D ones;
if $\beta < \beta_*$, 2D solitons carry more liquid than 3D ones.
Analysis of experimental data shows that the average separation
distance $L_x$ for 3D waves is much smaller than for 2D, in other
words, 3D waves are ``better packed'' in the same domain than 2D
waves. As a result $\alpha_{2D}/\alpha_{3D}$ is small which
together with the fact that $J_{2D}/J_{3D}$ is small (Eq. (1)
implies from (10) that $\beta_*$ is small as well). Hence, provided
that the average separation distance $L_{z}$ of 3D waves is not
very large, and the experiments show that it is not, 3D waves
carry more liquid than 2D ones.  Hence, both cases are possible,
depending on experimental conditions.

\section{Experiment}

\begin{figure}[p]
\centering\includegraphics[width=14cm]{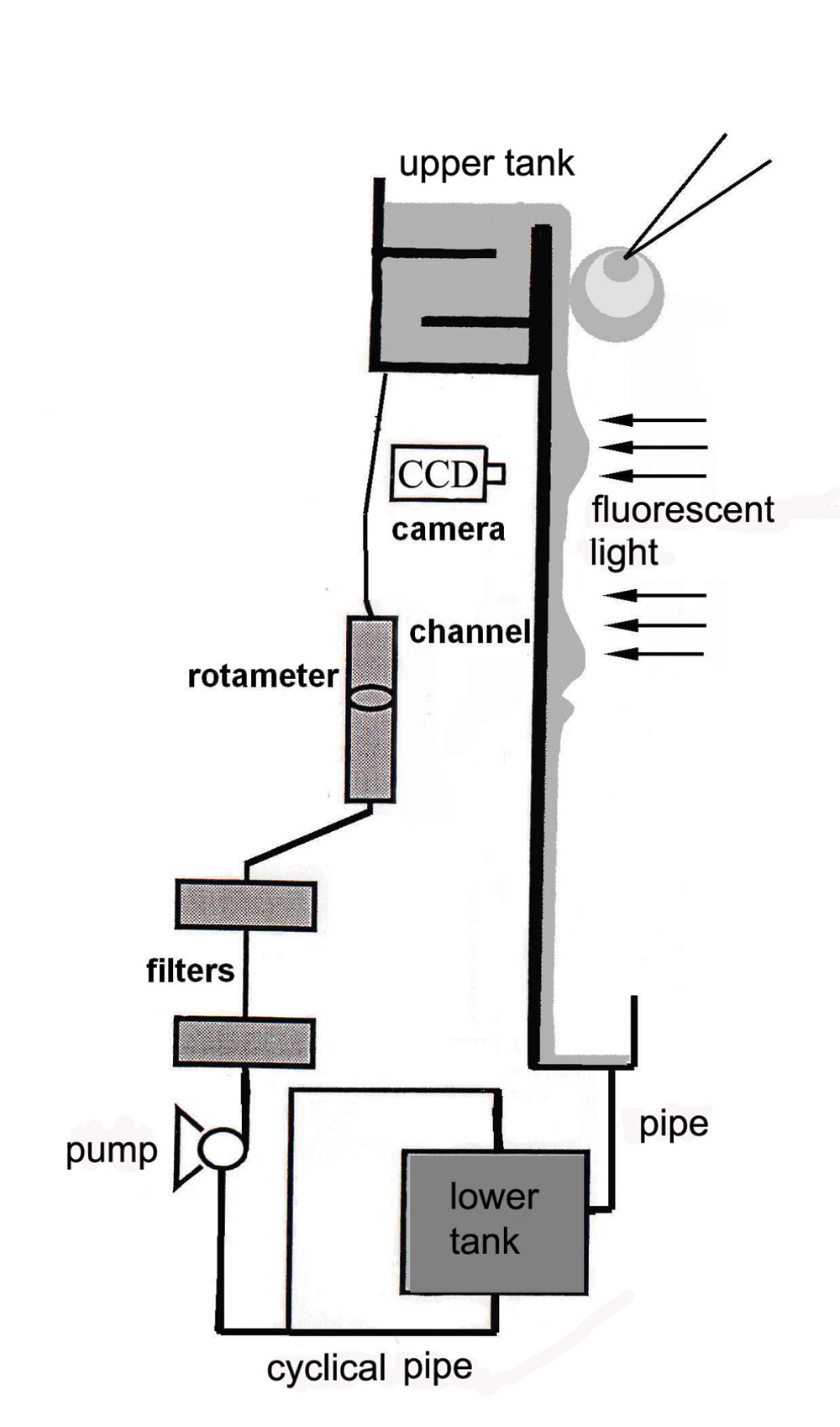}
\caption{Experimental setup. }\label{fig2}
\end{figure}

\emph{Experimental setup}.\\
We are going to
 to investigate the range of Reynolds numbers $\langle Re \rangle$ from 5 to 100. For
such Reynolds numbers, previous works have analyzed the flow of liquid
films falling down vertical tubes with diameters from 2 to 4~cm and
with water as test liquid. However, the typical soliton size in our setup
is 3 to 7~cm in the spanwise direction, so such tube
diameters are too small.   We employ an experimental apparatus
  with  a planar vertical channel,  25~cm
long and 15~cm wide and made of a special mirrored glass, see
Fig.~\ref{fig2}.

A digital pump is located in the lower part of the setup since
the flow is sensitive to external vibrations. Liquid is pumped
through a rubber pipe to a supply tank of size 25~cm
$\times$ 15~cm $\times$ 10~cm. In the upper part of the tank, there
is a feeding device of overflow type and the liquid flows
out of it through a special polymer net onto the vertical test
surface. The net serves to decrease the tangential velocity of the
liquid. From the test section, the liquid flows to the discharge
tank. The setup forms a closed loop, such that the liquid returns
from the discharge tank to the pump.

Distilled water is used as working fluid. The temperature is
kept at $15\pm0.5^\circ$C, so that the viscosity is
$\nu=1.14\times10^{-6}$~$\mbox{m}^2/\mbox{s}$, the ratio of
surface tension to density is
$\sigma/\rho=74\times10^{-6}$~$\mbox{m}^3/\mbox{s}^2$ and the
Kapitsa number is $\gamma=2900$. The Reynolds number at the inlet,
$\langle Re \rangle$, is measured as the discharge flow per unit width
divided by the viscosity $\nu$. The discharge volumetric flow rate of the pump is
varied from 7 to 150~$\mbox{cm}^3/\mbox{s}$ and hence the
corresponding range of Reynolds numbers is $4.5 < \langle Re
\rangle < 100$.

It is well known that the main difficulty with an overflow feed
in falling film experiments is the formation of a uniform flow at
the inlet and further downstream. With water as working liquid,
the film can be very thin. As a result, a small disturbance of the
overflow at the inlet leads to the liquid breaking up into drops
and rivulets. To remedy this effect, we take the following
measures: a)~the upper edge of the channel is aligned strictly
perpendicular to the gravity field; b)~the channel has a small
width, 15~cm, as mentioned earlier; c)~a polymer net is installed
at the inlet to decrease the tangential component of velocity;
d)~in the inlet region, we engrave ten small parallel ditches
streamwise, each 15~mm long and of triangular cross-section. These
measures significantly minimize the formation of drops and
rivulets. Still, at small Reynolds numbers, $\langle Re \rangle <
3 - 4$, we are not able to create a plane-parallel substrate.
Finally, polished steel side walls are used to minimize the
side-wall effect associated with flows in rectangular channels.

\emph{Methods to measure wave characteristics}.\\
 To measure 3D wave profiles and their
characteristics, we use the fluorescent imaging method by Liu and
Gollub  $^{\cite{L4}}$. The method was also successfully used by
Vlachogiannis and Bontozoglou $^{\cite{V1}}$. A fluorescent dye
with a low concentration (about 100 - 150 ppm), with chemical
formula $C_{20}H_{10}Na_{2}O_{5}$, is added to the fluid. Such a
small concentration does not affect the liquid physical properties
$^{\cite{L4}}$.  Eight ultraviolet lamps are arranged over the
channel surface to illuminate it from above and a chamber is
arranged on the opposite side of the channel. The absolute
accuracy of the method is 8 to 10 $\mu m$, hence, the largest
possible relative error varies from 6 to 8 $\% $.

The local film thickness can be obtained from the signal by
careful calibration. To perform the calibration of film thickness
versus light intensity, a part of the layer near the
distributor  is used. It remains flat up to the inception point which
varied from 7 to 25 cm. In the measurements, we avoid a small
portion of the channel just after the inlet to let the flow
establish a steady semi-parabolic profile (around ten to twenty
film thicknesses). The flat portion is used for calibration along
with the known discharge of the pump. The obtained dependence is
shown in Table 1. Between these points, a linear interpolation is
used,
$$I(x,z,t)=K*I_0(x,z)h(x,z,t)+I_1(x,z)$$
where $K$ is constant, and $I_0(x,z)$ and $I_1(x,z)$
depend on the location because of the non-uniformity of the
ultraviolet light field.

  The digital analogue of the measured physical signal of the surface is
recorded. Photographs are taken with a high-resolution
Panasonic NV-GS75 digital camera. The data are recorded on a
digital carrier and further processed using a computer with a
spatial resolution of 720 x 540. In order to measure the wave speed, a
series of frames of the wave are captured at $1/30-th$ of a
second apart.

Typical examples of shadow images of waves, their fluorescence
images and cross-sections for forced 2D periodic waves with
$\langle Re \rangle=30$ and frequency $f=15.0$~Hz and for a 3D
signal with $Re=6.8$ are shown in Fig. 4.

To measure the wave's amplitude by fluorescence
imaging, we use a simple contact method in which a steel
needle is positioned in the lower part of the channel. The needle
could move back and forth by means of an adjusting micrometre
screw. The flat surface of the substrate is used as a reference
point. Both electrical and visual methods are employed. In the
first case, the contact is registered by closing the electrical
circuit between the needle and contacting liquid. A very small
amount of sodium chloride which does not affect the hydrodynamical
properties of the working liquid is added to it to make it
conductive. The circuit includes a 4-volt battery and an
ampere-meter. To measure the maximum amplitude of the wave, we need
only one contact with the surface after which the experiment is
stopped. In the visual method, the circuit is switched off and on by
the top of the needle where a small coloured particle is positioned. When
the surface is touched by the needle, the particle is taken by the
flow. A microscope is employed to attach the particle and to
decide if the particle is taken by the flow or not.

\emph{Methods to investigate 2D-3D transition and 3D solitary waves}.\\
In order to form  2D solitary waves, localized periodic pulsations in
flow rate are superimposed on the main flow $^{\cite{A5,L2}}$.
The resultant ``artificial'' waves suppress the natural
perturbations and a rather regular train of 2D solitary waves
travelling downstream is generated. In order to induce a 3D
instability and the process of disintegration of 2D solitary waves
into 3D localized coherent structures, we impose artificial
perturbations onto the 2D solitons. For this purpose special
needles are placed in the transverse direction at the inlet, with
a separation distance $\tilde{l}$ from each other, as was done
in $^{\cite{P1}}$. We perform experiments with  $\tilde{l}$  from
$10$~mm to 35~mm in steps of $2.5$~mm.

Let us now discuss how to create 3D solitary waves. Previous
falling film experiments show that 3D solitons can be seen
after a long and complex downstream evolution $^{\cite{A4,A2}}$
(see also Fig.~1, Part~I). The key idea of our work is to bypass
all intermediate stages of evolution and deposit a single drop
with a fixed mass onto the smooth film surface. The advantage of
our method is that we deposit a drop with a mass close to that
of the theoretical 3D solitary wave. We  refer to this method as
the ``raindrop'' method. It was prompted literally by the rain on
a car windscreen during a traffic jam. This method provides a
simple, easy to implement and efficient way to successfully create
individual 3D solitons and to measure their parameters as a
function of the substrate Reynolds number.

 The depositing of drops is done with
a special small Finn pipette with adjustable volume. We have
found this device to be the most appropriate for our purposes. To
cover the range of soliton volumes obtained in Part~II,
reproduced here for the ease of presentation in Table~2, we employ
three Finn pipettes with the following drop volumes:
0.5~--~10~$\mu$L in increments of $0.1$~$\mu$L; 5~--~40~$\mu$L in
increments of $0.5$~$\mu$L and 40~--~200 $\mu$L in increments of
1~$\mu$L. The smallest diameter of the liquid drop is 2~mm and the
largest is 7~mm. The calculated dependence of soliton mass on
Reynolds number in Table~2 is used to estimate the mass of
the liquid drop. More specifically, the mass of the liquid drop is
taken as $1.1$ to $1.2$ times that of the equilibrium mass. Having a
liquid drop mass close to the steady 3D soliton mass results in
rapid formation of a 3D soliton.

In the experiment, the pipette is attached to the upper part of the
channel such that its distance and angle to the surface can be
changed. The most painstaking procedure is to adjust the pipette
in such a way that the moment the required drop size is achieved
the drop touches the film surface and is subsequently involved in
film movement. A schematic picture is given in
Fig.~\ref{fig8}. We have two degrees of freedom to approach the
film surface, distance and angle to the surface, which are
adjusted independently for all the flow rates used. During
adjustment, the image of the drop is magnified by a strong lens
located near the pipette. The distance and angle to the
surface are changed by employing adjusting micrometre screws.
Finally, a table of angles and distances as a function of flow
rate is obtained. After the drop touches the surface, it is
subsequently involved in film movement and is rapidly
transformed into a stationary localized wave.

The important condition in this method is to have a flat film surface
for the deposited drop. Recall that the film remains flat from the
inlet up to the inception point. The length of this region is
varied from 7 to 25 cm. The drop is deposited onto this part of the
flow.

 We found that it takes from 5 to 10~cm for the
$\Lambda$\nobreakdash-wave to reach a steady state --- the
unstable ripples at $\langle Re \rangle < 15$ are too weak to affect
the formation of the $\Lambda$\nobreakdash-wave in the upper part
of the channel. In this region of Reynolds numbers, the upper part
of the flow can be assumed to be practically flat. At $\langle Re
\rangle \simeq 20$, room disturbances have a sufficiently large
growth rate to turn into natural waves; although they continuously
disturb the configuration of a $\Lambda$\nobreakdash-wave sliding
on a flat substrate, they are not able to destroy it: the
$\Lambda$\nobreakdash-structure also grows with increasing
Reynolds number and is sufficiently robust to withstand the
natural waves.

The details regarding the apparatus and the experimental
techniques can be found elsewhere $^{\cite{S6}}$.

\emph{Numerical experiment}.\\
To complete our investigation, we perform numerical modelling of the
non-stationary process of
$\Lambda$\nobreakdash-soliton creation from a drop as done in the
physical experiments. As in $^{\cite{D1,D2}}$, we use the
following Kapitsa-Shkadov system of three nonlinear PDEs:

$$
 \frac{\partial q}{\partial t}+\frac{6}{5}\frac{\partial }{\partial x}\frac{q^2}{h}+\frac{6}{5}\frac{\partial }{\partial z}\frac{q p}{h} =
 \frac{1}{5 \delta}(h\frac{\partial}{\partial x}\nabla^2h +h-\frac{q}{h^2}),
$$
\begin{equation}\label{eq112Dim}
\frac{\partial p}{\partial t}+\frac{6}{5}\frac{\partial }{\partial
x}\frac{q p}{h}+\frac{6}{5}\frac{\partial }{\partial z}\frac{
p^2}{h} = \frac{1}{5 \delta}(h\frac{\partial}{\partial z}\nabla^2h
-\frac{p}{h^2}),
\end{equation}
$$
\frac{\partial h}{\partial t}+\frac{\partial q}{\partial
x}+\frac{\partial p}{\partial z}=0,
$$
where $\nabla^2={\partial^2}/ {\partial x^2}+{\partial^2} /
{\partial x^2}$ is the Laplace operator on the $(x,z)$ plane. All
the values are dimensionless and incorporate the liquid density
$\rho$, thickness ${h}_0$ and averaged velocity ${u}_0=(gh_0^2)/(3
\nu)$ on the flat substrate. The further transformation,
$$
x \to \kappa x,\quad z \to \kappa z, \quad t \to \kappa t, \quad v
\to v/ \kappa, \quad \kappa=\frac{\gamma^{1/3}}{3^{2/9}Re^{2/9}}
$$
allows to absorb two parameters, the substrate Reynolds number
$Re$ and Kapitsa number $\gamma=\sigma \rho^{-1} \nu ^{-4/3} g
^{1/3}$, into a single parameter,
$$
\delta=\frac{Re^{11/9}}{3^{7/9}5 \gamma^{1/3}}.
$$
Here, $\sigma$ is the surface tension, $\rho$ and $\nu$ are the
liquid density and viscosity,  and $g$ is the acceleration of gravity.

The numerical method of $^{\cite{D1}}$ is used, which is an
extension to 3D of the 2D scheme developed in the studies of
$^{\cite{C7}}$. For a localized 3D signal, the calculations are
performed in a frame moving with the velocity of the signal. The
computational domain is $2L_x$ x $2L_z$. A sufficiently large size
for the domain allows the final $\Lambda$\nobreakdash-wave
solution to be free from ``side-wall'' effects, or, equivalently,
the localized wave has sufficient space to decay almost to zero at
the boundaries of the domain. We impose boundary conditions at
both ends of the domain,
$$
 x=-L_{x}:\qquad h=q=1,
$$
$$
\quad \qquad x=-+L_{x}:\qquad h=q=1,\quad p=0,
$$
\begin{equation}\label{eq114Dim}
z=\pm L_{z}: h(x,-L_{z},t)=h(x,L_{z},t),
\end{equation}
$$
\qquad \qquad q(x,-L_{z},t)=q(x,L_{z},t),
$$
$$
\qquad \qquad p(x,-L_{z},t)=p(x,L_{z},t),
$$
  The initial conditions are taken
in the form,
\begin{equation}\label{eq113Dim}
  t=0: \qquad h=1+Ae^{-b(x^2+z^2)}, \quad q=1, \quad p=0
\end{equation}
which simulates a drop on the film surface with parameters $A$ and
$b$ describing the amplitude and spreading of the drop,
respectively.

\section{Results}
\emph{Visual observations of natural waves and their transitions}.\\
First, we consider experiments with waves originated from
natural room disturbances. A visual examination of the surface
shows that it is absolutely flat at $\langle Re \rangle < 4$~--~5,
confirming previous observations, $\langle Re^{(1)} \rangle
\approx 4$~--~5. Note that even at $\langle Re \rangle=7$,
the manifestation of waves is very weak, see Fig.~\ref{fig4}(a).
For larger Reynolds numbers, the surface is covered with waves.
At any $\langle Re \rangle$, experiments show the presence of a flat
region near the inlet up to the inception point.  At $\langle Re
\rangle \ge 7$, the points of wave inception are located on a line
at the following distance from the inlet: $\langle Re \rangle
=7$~--~25~cm; 10~--~15~$cm$; 15~--~10~$cm$; 20~--~7~$cm$. This
distance  is a little larger than cited in other works (see
$^{\cite{A4}}$) and $^{\cite{C7}}$). We attribute the difference
to the small width of the channel and wall effects.

For Reynonds numbers $\langle Re \rangle$ from 7 to 15 ~--~
20, the waves are two-dimensional and periodic. For Reynonds
numbers from 15~--~20 up to 40~--~50, the waves maintain the basic
structure of 2D solitons but are disturbed with some 3D
modulation, either stationary or pulsating. Localized 3D waves
can be seen in this range of Reynolds numbers but they have
small amplitudes and, what is more important, they have a short
lifetime and disappear rapidly downstream.

The 2D--3D transition occurs approximately at $\langle Re^{(2)}
\rangle \approx 40$~--~50, see Fig.~\ref{fig4}(b).  The 3D
modulation now is not stationary any more, but grows
downstream with eventual destruction of the 2D wave and appearance
of steady 3D solitons. We note that for $\langle Re \rangle >
\langle Re^{(2)} \rangle$, the coalescence of neighbouring 2D waves
increases the 3D modulation and accelerates the destruction
process. A typical well-developed 3D regime is shown in
Fig.~\ref{fig4}(c).  The following interesting visual observation is made
regarding the ST: if we look at a fixed point on the channel, thus
taking the Eulerian viewpoint, the evolution appears to be purely
chaotic. But if we follow the waves with an appropriate speed,
thus taking the Lagrangian viewpoint, we can decipher interacting
deterministic localized structures, which look like
``quasi-particles''. In other words,  the ST can be considered to be
a chaotic interaction of quasi-particles which are strongly
internally coupled and continuously interact with their neighbours.
 An example of the chaotic wave motion in this region is
shown in Fig.~\ref{fig7} for $\langle Re \rangle = 60$. Several
typical patterns can be identified: 1~--- 2D soliton with 3D
modulation prior to being destroyed downstream; 2~---
Well-developed 3D solitons interacting through their overlapping
tails; 3~--- Complex coalescence of several 3D solitons.

The 3D localized coherent structures slide ``rapidly'' on a
``slowly'' moving thin sublayer with Reynolds number $Re$. In the
3D wave regime, most of the liquid is concentrated in the 3D
structures and only a small fraction of it is carried by the thin
substrate. As we pointed out in the previous section, for our
range of Reynolds numbers we take this fraction as $\sim 10$ . As
a consequence, the above range of inlet Reynolds numbers for
3D wave regimes translates into $4 < Re < 40$.

\emph{Forced waves and comparison with our theory}.\\
The rather irregular picture of natural waves can be improved by
imposing artificial periodic pulsations in flow rate
on the main flow . Further downstream, the pulsations are convected into
regular 2D periodic waves, at low pulsation frequency the
periodic waves turn into 2D solitary waves. The initially 2D waves
are disturbed in the transverse direction by a system of needles
placed with a separation $\tilde{l}$ and the
regular structure of waves is immediately violated. At some
distance from the inlet, 3D modulations appear to be superimposed
on our 2D solitons. These modulations grow downstream and
eventually destroy the 2D solitons, and 3D localized structures
appear at some distance $\tilde{L}$ from the inlet. A typical
snapshot of such an evolution is shown in Fig.~\ref{fig5}, $\langle
Re \rangle = 10$, $\tilde{l}=25$~mm.

We perform experiments with different $\tilde{l}$, from $10$~mm to
35~mm, in steps of $2.5$~mm. For sufficiently short separation
between the needles, more specifically, $\tilde{l} \leq 10$~mm,
there is no sustained reaction and the 3D perturbations decay
downstream. At larger $\tilde{l}$, wave disintegration into
localized 3D coherent structures at a distance $\tilde{L}$ is
observed. From one experimental run to another, this distance
varies with an amplitude of approximately 0.5~cm. Our data are
averaged over 20 to~40 points to obtain $\tilde{L}$. The results
of our observations for $\tilde{L}$ are presented as a function of
the distance $\tilde{l}$ for $\langle Re \rangle = 10$ in
Fig.~\ref{fig6}. The dependence has a pronounced minimum at
$\tilde{l}=2$ cm which is in good agreement with our theoretical
prediction $^{\cite{D1}}$ of the most dangerous wavelength,
$\tilde{l}_m=1.42 Re^{1/9} =1.83$ cm, and with experimental
observations $^{\cite{P1}}$. Moreover, the downstream distance for
complete disintegration, $\tilde{L}$, is about 10 to 15~cm, which
also fits the theoretical prediction $8~-~15$ cm (Part I Fig. 10).

In the raindrop method, a 3D coherent structure is formed by
depositing a liquid drop onto the film surface. The drop is
subsequently involved in film movement. We emphasize that the
 drop's evolution results in a stationary 3D solitary wave only if the
 drop's mass is close enough to the equilibrium one. In this case,
 the drop is rapidly transformed into a stationary localized wave. It takes about 5 to
10~cm from the location where the drop touches the film surface for
a stationary structure to form. For a larger mass, approximately
as much as twice the equilibrium mass, a two-hump 3D solitary
wave is formed. This wave is unstable and decomposes further
downstream into two one-hump 3D solitary waves. If the mass of the
deposited drop is much smaller than the equilibrium mass,
that is approximately one half, then it never evolves into a stationary wave
but expands in a 3D wave packet.

A comparison of experimental and theoretical cross sections of
$\Lambda$\nobreakdash-solitons is given in
 Fig.~\ref{fig11}. There is a rather good
correspondence between the wave profiles, except for the capillary ripple
region in front of the wave --- theoretical oscillations are more
pronounced. This can be explained by insufficient accuracy in our
measurement in the region of capillary ripples, 8~--~10 $\mu m$.
Also, the theoretical trough behind the hump is deeper than the
experimental one, the difference becoming larger at large
$\delta$.

 The $\Lambda$\nobreakdash-wave profiles and other wave
characteristics can also be found in experimental works
$^{\cite{A7,A8}}$.  In $^{\cite{A8}}$, parameters for two solitary
waves are presented for $\gamma=404$: $Re=2.5, \quad
\delta=0.035$,
 and $Re=3.9, \quad \delta=0.061$. According to our
theory in Part II, at $\delta<0.054$ the
$\Lambda$\nobreakdash-soliton is unstable against the radiation mode,
in other words at small $\delta$, localized disturbances of the
plane-parallel part of the $\Lambda$-wave destroy the hump.
Indeed, the first experimental wave looks unsteady and shows
some indication of further disintegration. In Fig.~\ref{fig12},
we compare our theoretical (a) and experimental (b) wave profiles
and their cross-sections for the second case, $Re=3.9$,
$\delta=0.061$, $\gamma=404$. The experimental and theoretical
velocities are $209$ and $213\mbox{ mm}/\mbox{s}$, respectively.
For this $\delta$, the wave is stable and we can find good
quantitative agreement between our theory and experiment. Our
theory $^{\cite{D10}}$ also predicts the existence of two-hump 3D
solitons and their characteristics. Such solitons were found
experimentally in $^{\cite{A7}}$. The comparison of theoretical
and experimental data in Fig.~\ref{fig13} gives reasonably good
correspondence.

Experimental data for the wave speed is given in Table~3. For the
first two Reynolds numbers, we are not able to create a 3D soliton
because of its instability. Instead of a nonlinearly coupled
localized structure, we observe a spot of expanding chaotic waves
confirming that at $\delta<0.054$ ($Re<6$) 3D solitons are
unstable, in agreement with our theoretical predictions. We are not
able to sustain a stationary 3D solitary wave at $\delta \le
0.054$ because of a strong primary instability of the flat region.
However, at larger $\delta$, the radiation is convected away from
the 3D soliton and it is stable and robust. The dimensional values
from Table~3 are plotted in Fig.~\ref{fig14} along with the
theoretical ones. The agreement is really good especially taking
into account the simple way our experiment is performed.

Fig.~\ref{fig15} compares the experimentally observed wave
amplitudes as a function of Reynolds number with the theoretical
ones. Again, the agreement is rather good.

\emph{Numerical experiment --- comparison with physical results}.\\
  Typical numerical results of the
time-dependent evolution for $\delta=0.0908$ (for water $Re$ =
$9.28$) are shown in Fig.~\ref{fig16}. A localized `drop' with the
initial form described by (13) and with a mass 20\% larger than
the equilibrium 3D soliton mass is placed on the flat film
substrate at $t=0$. At $t=0.05$ the drop has a tendency to split
into two humps; however, provided this event is avoided, instead
of disintegration one nonlinear hump of the soliton is formed.
(The tendency to split into two humps cannot be arrested if we
take a drop with an initial mass at least twice the equilibrium
3D soliton mass.)  At time $t=1$ the signal acquires capillary
ripples (this moment is skipped in Fig.~\ref{fig16}), and at
$t=5$ the trough behind the head develops. At $t=12-20$ the 3D
soliton is practically formed. One can clearly see at this stage
the excess mass draining slowly to the back, leaving the
equilibrium 3D soliton behind it. The final structure is
characterized by capillary oscillations/ripples at its front along
with moustache-like or leg-like structures at the back and a long hollow trough between
the legs. The ditch is caused by a Bernoulli depression in the
moustaches and in fact the numerical experiments performed here
indicate that it develops simultaneously with the development of
the moustaches. A completely stationary $\Lambda$\nobreakdash-wave
running at speed $c \simeq 4.6$ is formed at $t=20$.

The results of these calculations are compared with our
experiments in Fig.~\ref{fig155}, where the wave amplitude
$h_{max}$ is presented as a function of the distance downstream.
The theoretical time is recalculated as the distance downstream
using the signal velocity and physical properties of water. The
dimensionless time of $12$ to $20$ corresponds to a dimensional
distance of $6$ to $10$~cm which is in a good agreement with the
physical experiments.

The calculations also confirm other experimental results: If the mass of
the {``drop''} is at least twice as large as the equilibrium 3D
soliton's mass, the drop disintegrates into two or several
solitons or into a two-hump 3D soliton.  If the drop's mass is
smaller than the equilibrium 3D soliton's mass, the structure
sucks up liquid from the substrate as it evolves and it takes two
to three times longer to reach the final
$\Lambda$\nobreakdash-wave's steady state. Moreover, starting
with a small initial mass, the evolution never results in a
stationary 3D wave but in an expanding and growing wave packet.

The evolution of different shapes of initial signal was
also investigated. The constants $A$ and $b$ in (13) are varied in
such a way that the volume of the initial signal remains the same.
In all cases, the main stages of the evolution appear to be
qualitatively the same. Interestingly, the time to reach a 3D
stationary structure appears to be almost independent of the form
of the initial signal. Furthermore, changing the initial mass of
the drop to $1.5$ times that of the equilibrium 3D solitary wave
also gives qualitatively the same stages of evolution and
finally results in a stationary 3D pulse with nearly the same time
of saturation $t\simeq 10$ to~$20$.

It is now well established that the interaction of 2D solitons
with smaller objects in front of them consists of a coalescence
process in which the solitons absorb the mass of the smaller
objects. This results in an increase in the speed and amplitude of
the solitons by an amount proportional to the mass of the smaller
objects. This coalescence process also takes place between 2D
solitons of different amplitudes and speeds: the faster soliton
catches up with the smaller one and absorbs its mass. The
coalescence event between 2D solitons has been observed
experimentally ($^{\cite{L2}}$) and has been described
theoretically  $^{\cite{C5}}$.

Fig.~\ref{fig17} presents some computational results of such an
interaction, but for 3D solitons. At $t=0$ we place in front of a
stationary 3D soliton a localized signal with a mass of
$\frac{1}{2}$ the mass of the soliton. At $t=5$, the soliton
absorbs the mass of the localized signal and retains it until
$t=25$, i.e. for a time interval of $\Delta t=20$. Recalculation
of this dimensionless time into a dimensional distance gives
roughly 10~cm. The coalescence event causes an instantaneous
acceleration and increase in speed/amplitude proportional to the
absorbed mass. The 3D soliton is now an ``excited'' soliton, i.e.
a soliton of larger amplitude than that corresponding to the
given $\delta$. For $t>25$, drainage of the excited soliton begins
and the excess mass leaves the wave. It is expelled from the
structure through its ``moustaches'' and eventually takes the form of
two small 3D solitons ($t=41$). These solitons suck up liquid from
the substrate and eventually grow into equilibrium 3D solitons. This
is the key mechanism for generating the staggered ``checkerboard''
pattern seen in many 3D falling film experiments. The beginning of
this process is seen at $t=45$ and $49$.

\section{Conclusions}

Through simple experiments, we have examined the instabilities and
transitions leading to the regime of ST. We have shown that if
2D solitary pulses are unstable to transverse perturbations, their
disintegration results in the formation of 3D
$\Lambda$\nobreakdash-solitons. For water, at sufficiently large
Reynolds numbers the most dangerous length is 2~cm; this result is
close to that obtained theoretically $^{\cite{D1}}$.

We have developed a new experimental technique, the ``rain-drop''
method, to investigate $\Lambda$\nobreakdash-solitons and their
characteristics. Wave shapes, amplitudes and velocities have been
obtained for a wide range of Reynolds number. Numerical
experiments modelling ``rain-drop'' formation have also been
performed. We have  established good agreement between the
theoretical and experimental characteristics and parameters of
$\Lambda$\nobreakdash-solitons.

An important next step in our investigation is the development of a
statistical theory of the ST. We also plan an experiment and a
direct numerical simulation of the 3D wave regime in a vertically
falling film to model in detail the downstream evolution of
natural waves, their interaction and statistics. In order to
extract the deterministic features of the 3D wave regime, the methods
of ``spatial averaging'' and ``proper orthogonal'' decomposition
used successfully in ordinary turbulence $^{\cite{A9}}$ might also
prove very useful in the falling film problem. We hope to
examine these and related problems in future studies.

\newpage
{\em Table 1}  {Calibration of thickness vs.
intensity for the flat film.}\\ \\
\vspace{20mm}
\begin{tabular}{|c|c|c|c|c|c|c|}
\hline $h$, $\mu m$ & $I$   \\
\hline
0   & 0  \\
\hline
141    & 24.8  \\
\hline
146   & 25.5   \\
\hline
152   & 26.3  \\
\hline
155   & 26.7  \\
\hline
163   & 27.8  \\
\hline
168   & 28.5  \\
\hline
173   & 29.2  \\
\hline
183  & 30.6  \\
\hline
191 & 31.6  \\
\hline
311 & 47.1  \\
\hline
503 & 69.5  \\
\hline
\end{tabular}

\newpage
{\em Table 2}  {Stretching coefficients $\kappa$,
 Reynolds number of substrate and
volume of equilibrium 3D-``drops'', dimensionless
and dimensional, as a function of $\delta$ for water, $\gamma=2900$.}\\ \\
\vspace{20mm}
\begin{tabular}{|c|c|c|c|c|c|c|}
\hline
$\delta$ & $\kappa$ & $Re$ &
$J_{3D}$ &
$\tilde{J_{3D}}, \mu L$  \\
\hline
0.005   & 11.48 & $0.861$ & 3.054 & 0.1378 \\
\hline
0.010    & 10.12 & $1.518$ & 4.748 & 0.2935 \\
\hline
0.015   & 9.403 & $2.116$ & 5.775 & 0.429  \\
\hline
0.020   & 8.924& $2.677$ & 7.327 & 0.621  \\
\hline
0.025   & 8.569 & $3.213$ & 8.739 & 0.819 \\
\hline
0.030   & 8.289 & $0.373$ & 10.483 & 1.068 \\
\hline
0.035   & 8.060 & $4.232$ & 12.720 & 1.389 \\
\hline
0.040   & 7.867 & $4.720$ & 15.491 & 1.798 \\
\hline
0.045  & 7.700 & $5.197$ & 19.374 & 2.373 \\
\hline
0.050 & 7.554 & $5.666$ & 24.518 & 3.150 \\
\hline
0.55 & 7.424 & $6.125$ & 31.920 & 4.283 \\
\hline
0.060 & 7.308 & $6.577$ & 40.862 & 5.704 \\
\hline
0.070 & 7.106 & $7.462$ & 63.692 & 9.536 \\
\hline
0.090 & 6.788 & $9.165$ & 101.52 & 17.039 \\
\hline
0.150 & 6.187 & $13.92$ & 223.56 & 47.330 \\
\hline
0.200 & 5.871 & $17.61$ & 325.26 & 78.481 \\
\hline
0.500 & 4.970 & $37.28$ & 935.46 & 343.31 \\
\hline
\end{tabular}

\newpage
\vspace{22mm} {\em Table 3}  {Pipette drop volume ${J}$, Reynolds
number $Re$, wave velocity ${c}$, and maximum thickness of
wave ${h_{max}}$,  as a function of supply tank volumetric discharge rate,
${Q}$ (water).} \\
\\
\begin{tabular}{|c|c|c|c|c|}
\hline ${Q}, cm^3/s$ & $Re $ & ${J},\mu L$ & ${c}, cm/s$
& ${h_{max}},mm$\\
\hline
7.50   & 4.38 & 2.00 & - & $ - $\\
\hline
9.00    & 5.26 & 3.00 & - & $ - $ \\
\hline
10.5   & 6.72 & 5.00 & 22.5 & $0.20$\\
\hline
12.0   & 7.02 & 9.00 & 24.5 & $0.28$\\
\hline
13.5   & 7.89 & 15.0 & 27.7 & $0.32$\\
\hline
15.0   & 8.77 & 20.0 & 31.1 & $0.35$\\
\hline
16.5   & 9.65 & 25.0 & 33.9 & $0.41$\\
\hline
19.5 & 11.40 & 40.0 & 37.5 & $0.39$\\
\hline
21.0   & 12.28 & 45.0 & 39.3 & $0.40$\\
\hline
22.5   & 13.16 & 60.0 & 42.6 & $0.43$\\
\hline
24.0   & 14.03 & 60.0 & 43.2 & $0.47$\\
\hline
24.0   & 14.03 & 70.0 & 42.1 & $0.48$\\
\hline
27.0   & 14.91 & 75.0 & 41.1 & $0.51$\\
\hline
27.0   & 15.79 & 80.0 & 46.0 & $0.47$\\
\hline
28.5   & 16.67 & 90.0 & 47.8 & $0.48$\\
\hline
30.0   & 17.54 & 100.0 & 48.9 & $0.50$\\
\hline

\end{tabular}

\newpage

\begin{figure}[p]
\centering
\includegraphics[width=16cm]{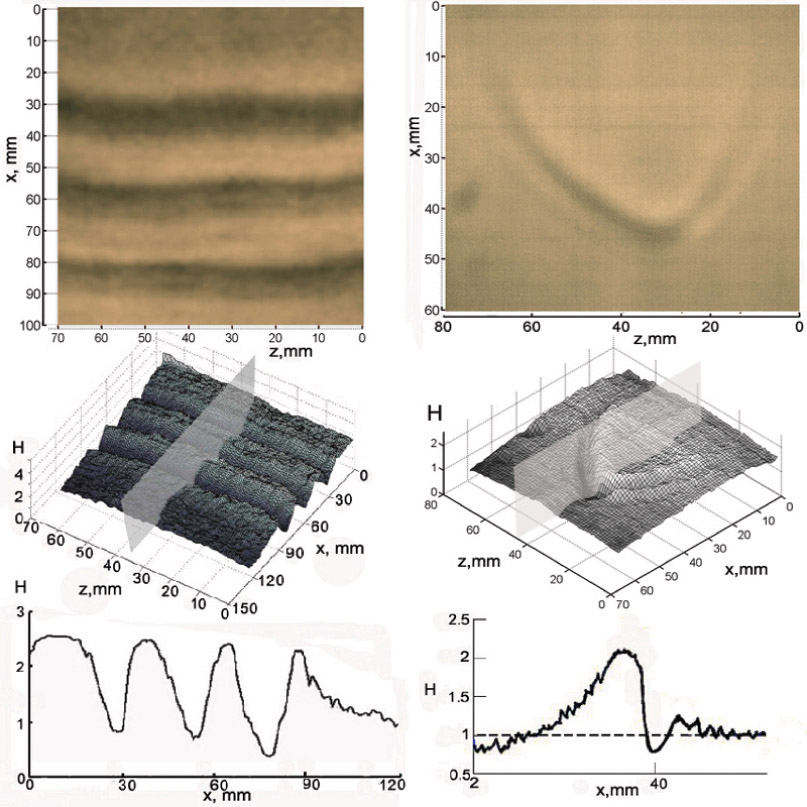}
\caption{Examples of shadow images of waves, fluorescence
images and cross sections: (a)Forced 2D periodic waves for $ Re
=30$ and frequency $f=15.0$ Hz. (b) 3D-soliton for $Re=7$.
}\label{fig3}
\end{figure}

\begin{figure}[p]
\centering
\includegraphics[width=7cm]{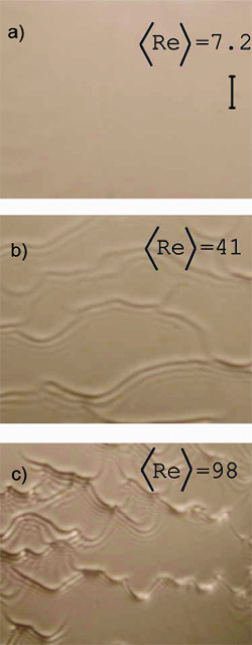}
\caption{Shadow images of waves generated by room disturbances.
The bar is 2 cm long. (a)$\langle Re \rangle=7$, the flow is
practically flat, (b)$\langle Re \rangle=41$, 2D-3D transition and
(c)$\langle Re \rangle=98$.}\label{fig4}
\end{figure}

\begin{figure}[p]
\centering
\includegraphics[width=11cm]{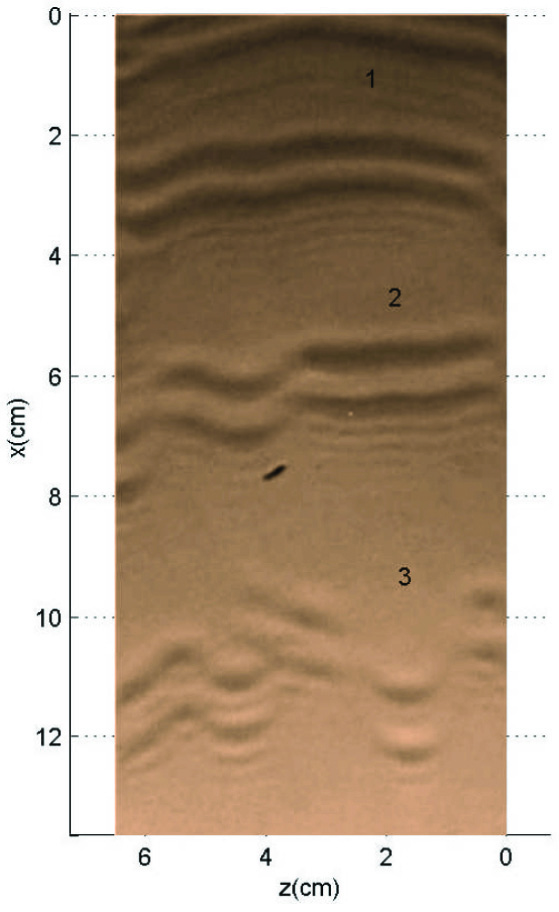}
\caption{Shadow image of periodically excited waves for $Re=10$.
The waves are disturbed in the spanwise direction with needles at
the inlet with separation $\tilde{l}=25$ mm. At point 1, the wave is
practically two-dimensional; at point 2 at a distance of 6 cm from
the inlet, one can clearly see the 3D modulation, at point 3 at a
distance $\tilde{L}=11$ cm from the inlet, the original waves are
completely destroyed.}\label{fig5}
\end{figure}

\begin{figure}[p]
\centering
\includegraphics[width=15cm]{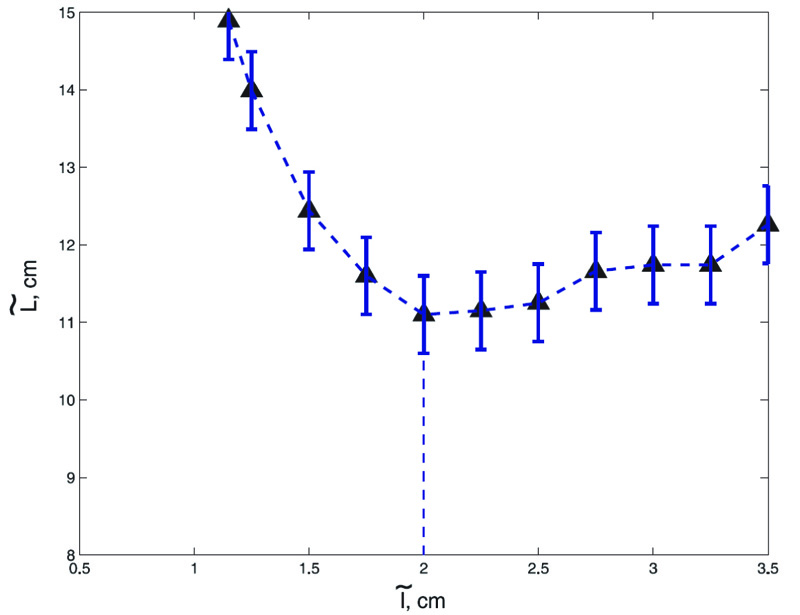}
\caption{The locus of the 2D-3D transition for $\tilde{L}$ as a
function of the length of the transverse perturbation $\tilde{l}$
for $Re=10$. There is a pronounced minimum at $\tilde{l}=2$
cm.}\label{fig6}
\end{figure}

\begin{figure}[p]
\centering
\includegraphics[width=10cm]{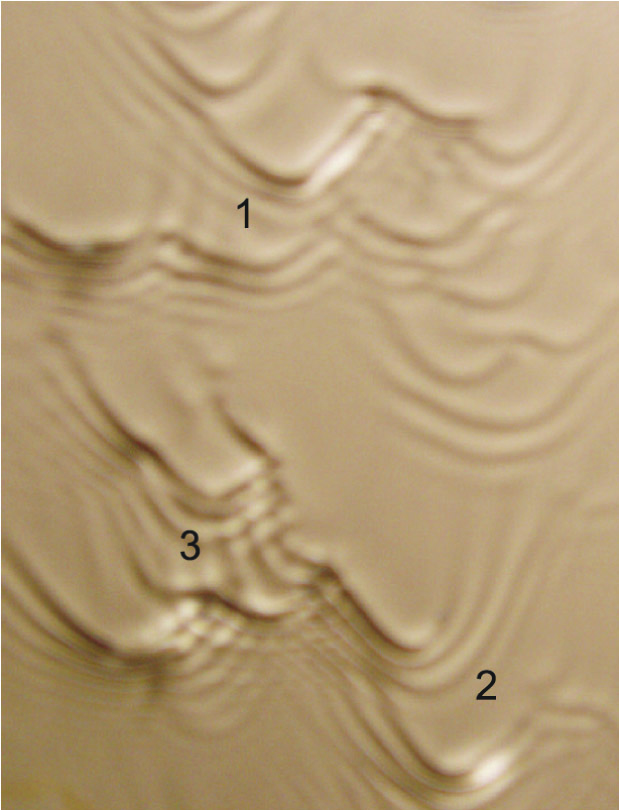}
\caption{Typical example of the chaotic wave motion observed for
$\langle Re \rangle = 60$. This section is located between 14 cm
and 24 cm from the inlet. There are several typical wave patterns:
1 - 2D soliton with 3D modulation ready to be destroyed
downstream; 2 - Well-developed 3D solitons interacting through
their tails; 3 - Complex collision of several 3D
solitons.}\label{fig7}
\end{figure}

\begin{figure}[p]
\centering
\includegraphics[width=9cm]{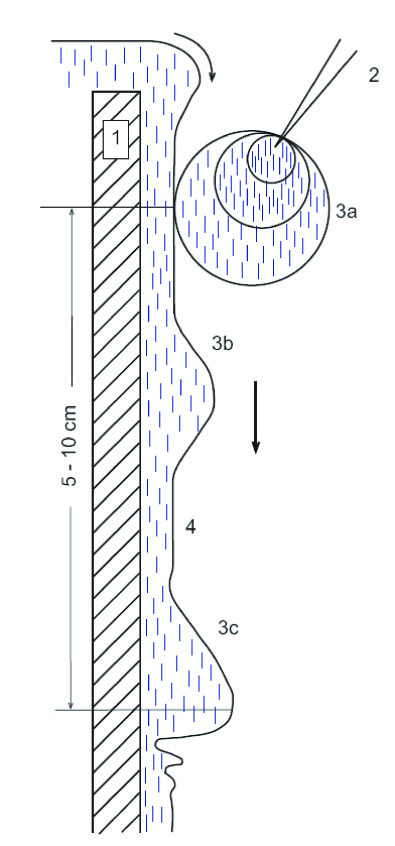}
\caption{``Rain-drop'' method of exciting 3D soliton with the
mass of liquid in the pipette corresponding to the equilibrium
3D soliton mass. 1 - Test glass channel; 2 - Pipette with
adjustable drop-volume; 3 - Successive instants of drop evolution,
namely, 3a - drop is gradually increasing in size and finally
touches the liquid surface and becomes involved in film flow, 3b -
drop spreads over and turns into an evolving localized
signal downstream which eventually 3c - becomes a stationary running
$\Lambda$ wave; 4 - liquid layer flowing down the channel.
}\label{fig8}
\end{figure}

\begin{figure}[p]
\centering
\includegraphics[width=17cm]{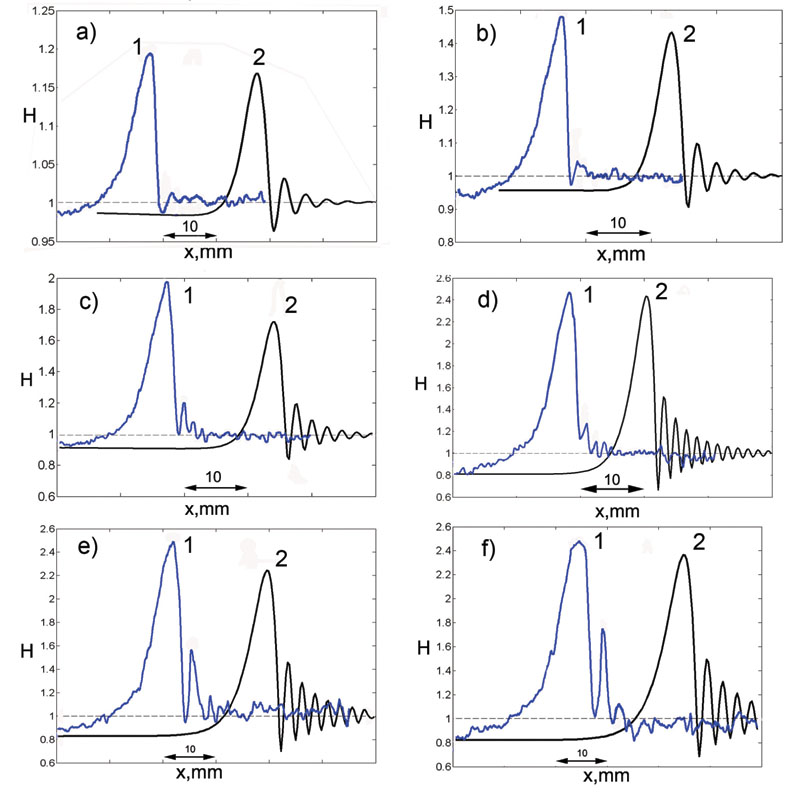}
\caption{Experimental (1) and theoretical (2) cross sections,
(a)$Re=7$, (b)$Re=8.8$, (c)$Re=9.7$, (d)$Re=13.2$, (e) $Re=15$,
(f) $Re=17.7$.}\label{fig11}
\end{figure}

\begin{figure}[p]
\centering
\includegraphics[width=12cm]{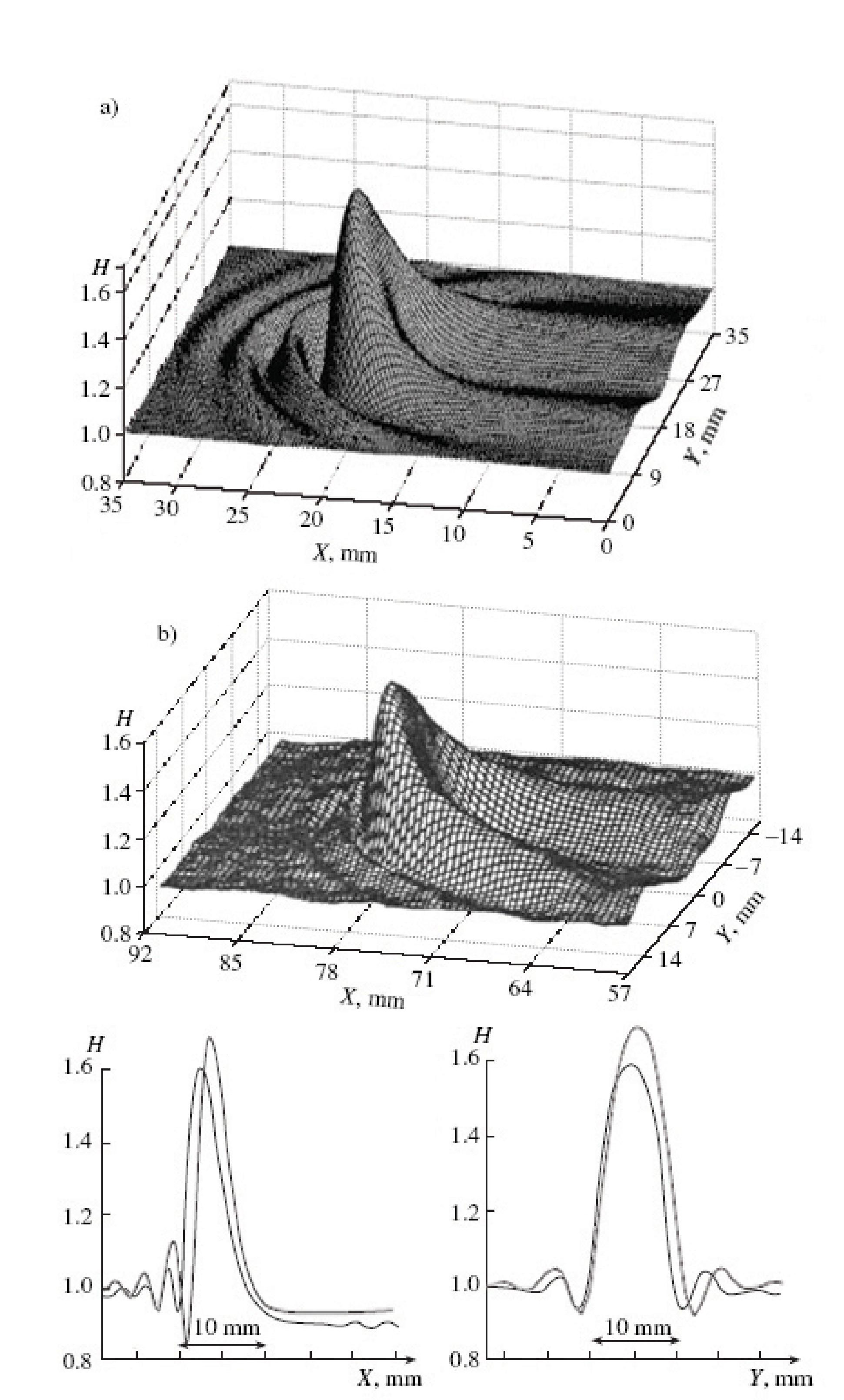}
\caption{Comparison of our theoretical (a) and experimental
(b) (see  Alekseenko et al. \cite{A8}, Fig. 4) profiles of
$\Lambda$ solitons and their cross sections for $Re=3.9, \
\delta=0.061,\ \gamma=404$. Experimental wave velocity is $209\
mm/s$ and theoretical velocity is $213\ mm/s$.}\label{fig12}
\end{figure}

\begin{figure}[p]
\centering
\includegraphics[width=12cm]{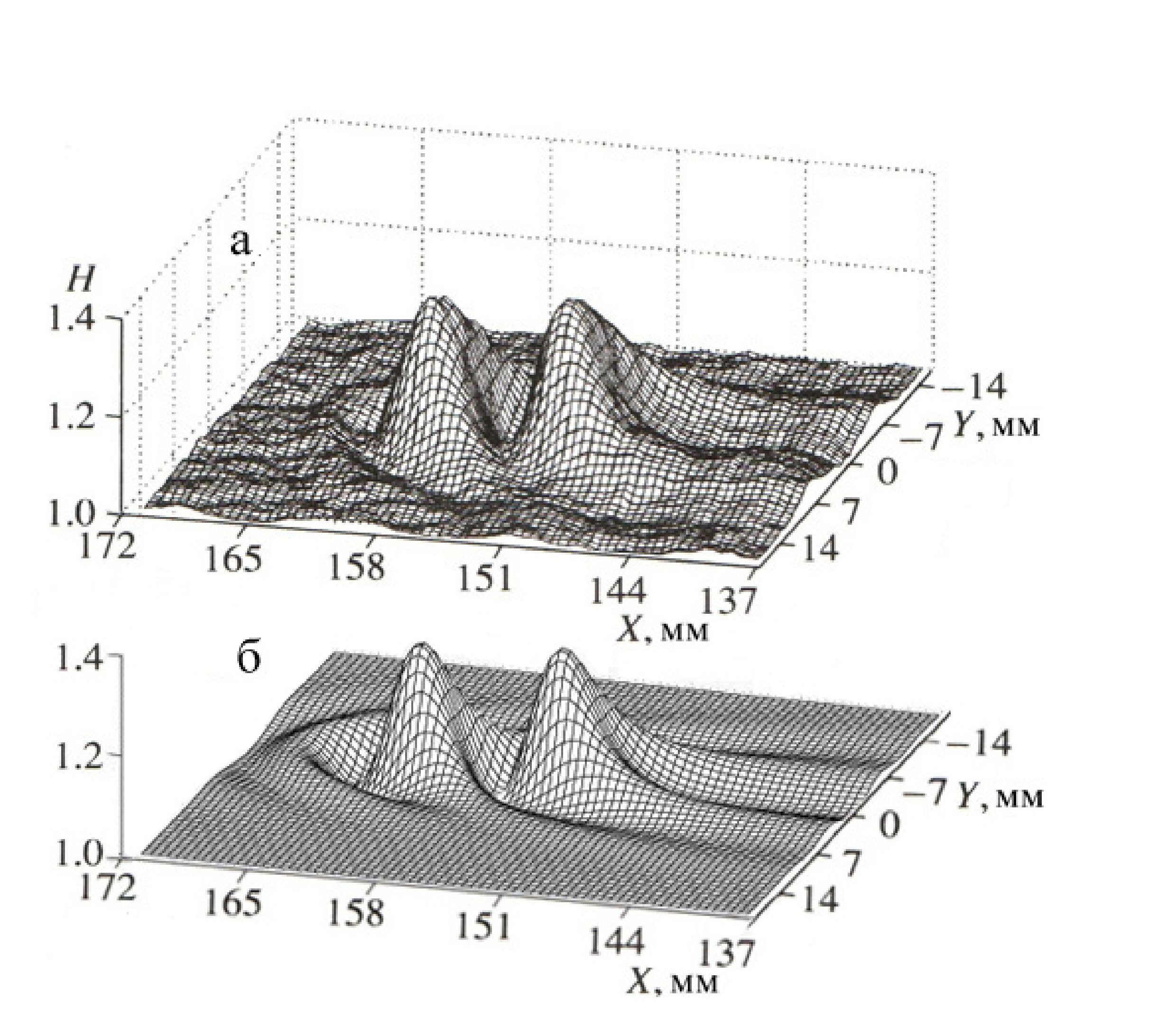}
\caption{Comparison of our theoretical (a) and experimental
(b)(see  Alekseenko et al. \cite{A7}, Fig. 3) profiles of two-hump
3D solitons  for $Re=2.2, \ \delta=0.03, \ \gamma=404$.
Experimental wave velocity is $102\ mm/s$ and theoretical velocity
is $113\ mm/s$.}\label{fig13}
\end{figure}

\begin{figure}[p]
\centering
\includegraphics[width=15cm]{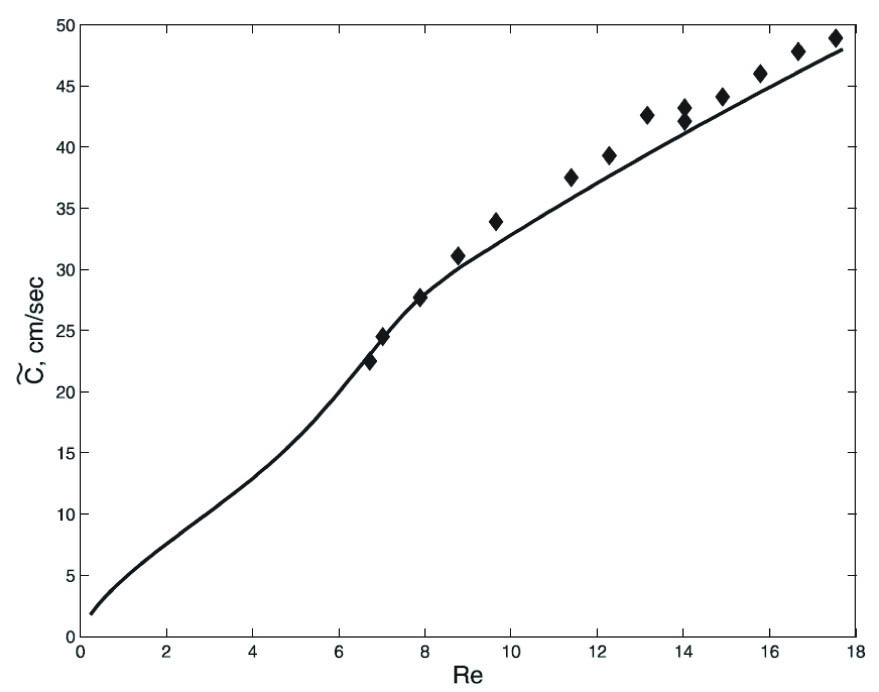}
\caption{ Phase velocities of $\Lambda$-solitons as a function of
$Re$. The diamonds represent our experimental
results.}\label{fig14}
\end{figure}

\begin{figure}[p]
\centering
\includegraphics[width=15cm]{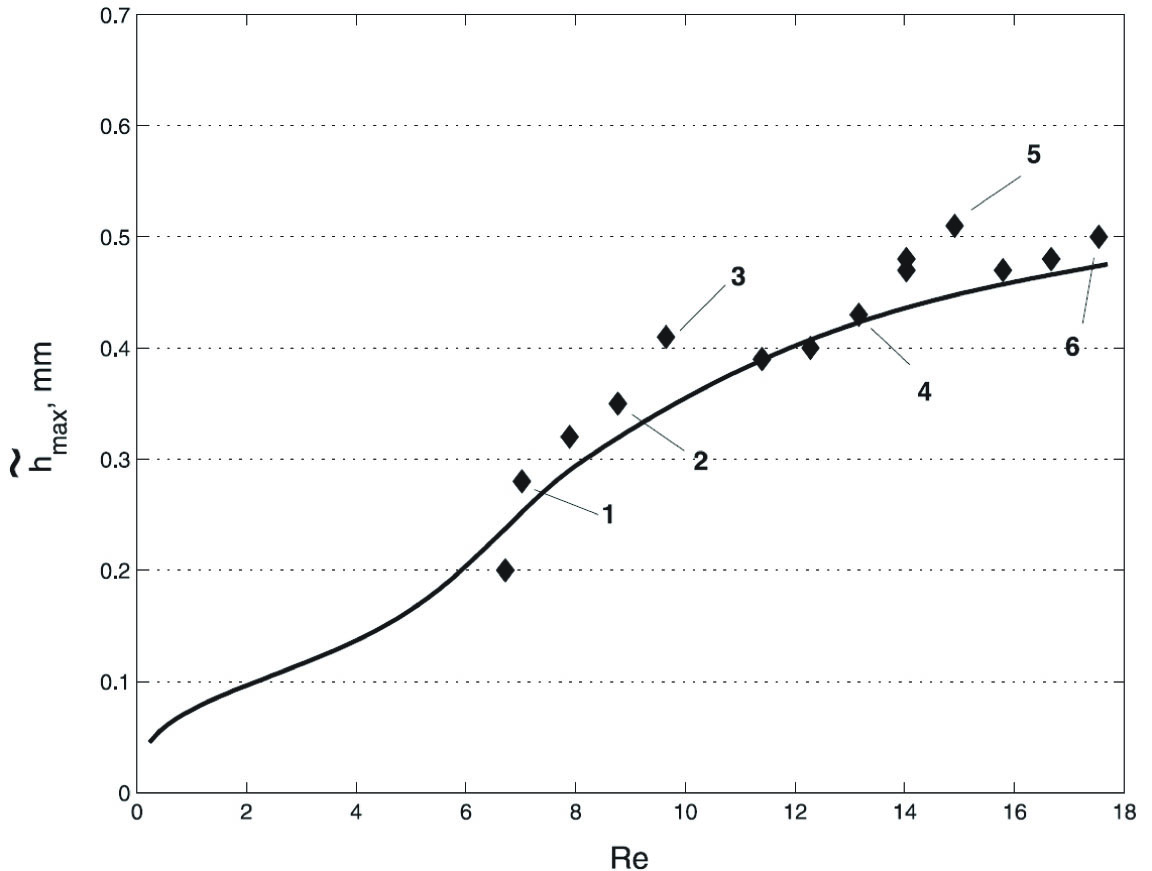}
\caption{ Maximum amplitudes of $\Lambda$-solitons as a function
of $Re$. The diamonds represent our experimental
results.}\label{fig15}
\end{figure}

\begin{figure}[p]
\centering
\includegraphics[width=17cm]{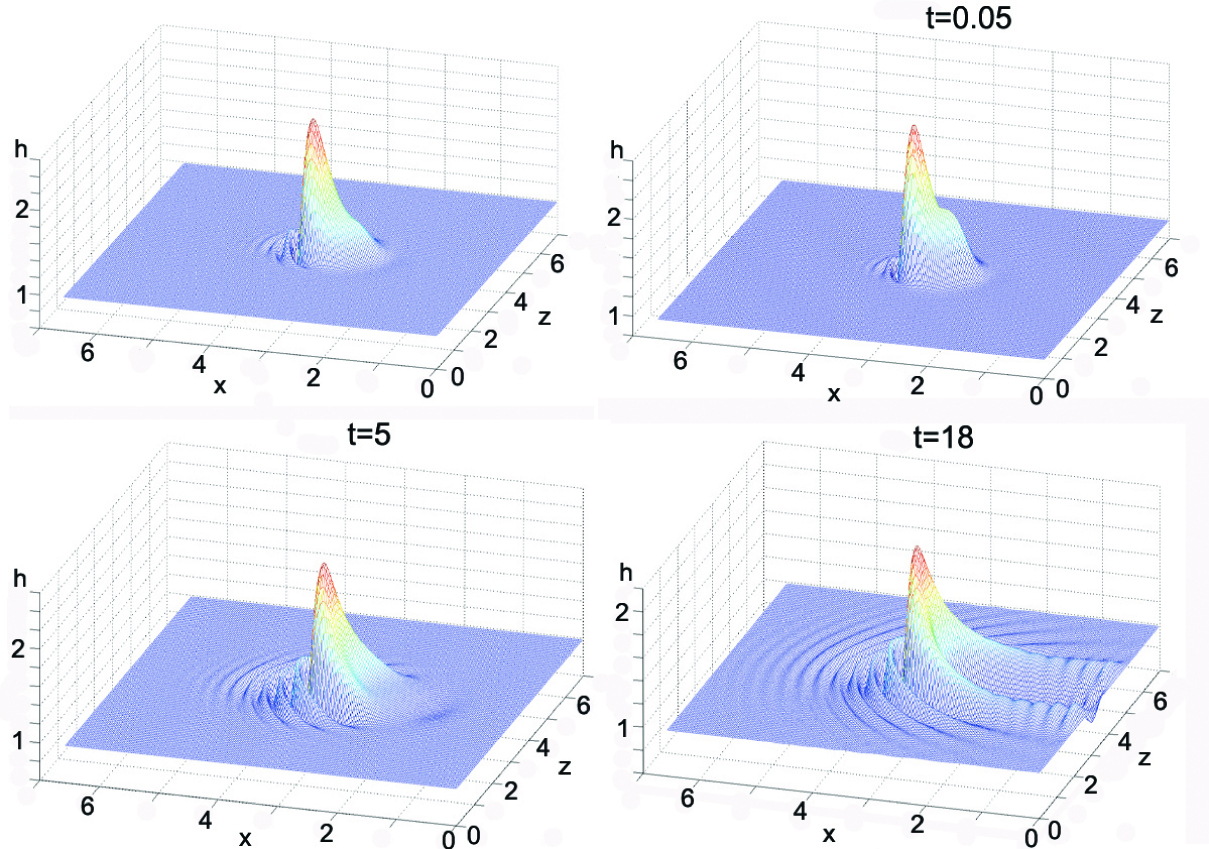}
\caption{ Formation  of  a $\Lambda$ soliton from a localized
signal, or ``drop'', for $\delta=0.0908$ at different times. At
$t=0.05$ the drop has a tendency to disintegrate. At $t=1.0$
the capillary ripples form (not shown). At $t=5.0$ the trough
behind the head forms. At $t=12.0$ we have reached a nearly steady
3D coherent structure. Finally, at $t=18.0$ the 3D structure takes
the shape of a $\Lambda$-soliton.}\label{fig16}
\end{figure}

\begin{figure}[p]
\centering
\includegraphics[width=15cm]{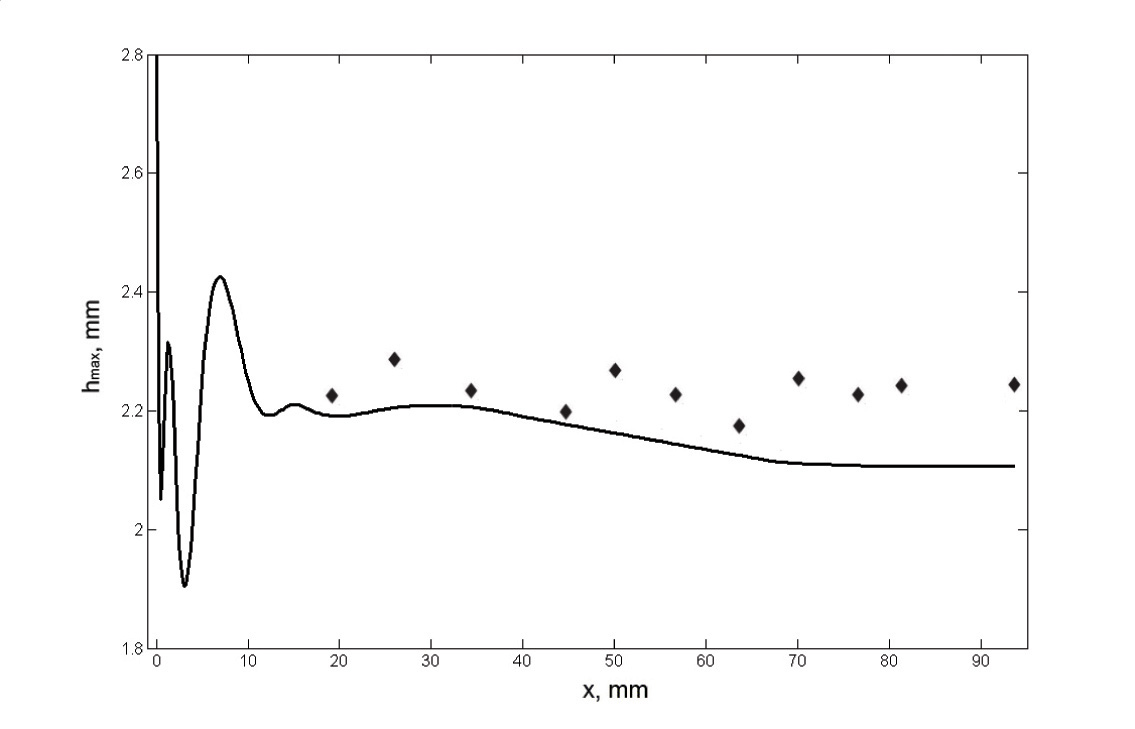}
\caption{ Evolution of the dimensionless amplitude $h_{max}$
downsteam, $\delta=0.0908$, $Re=9.28$ (water). The line stands for
the calculations, the diamonds for our experimental
results.}\label{fig155}
\end{figure}

\begin{figure}[p]
\centering
\includegraphics[width=17cm]{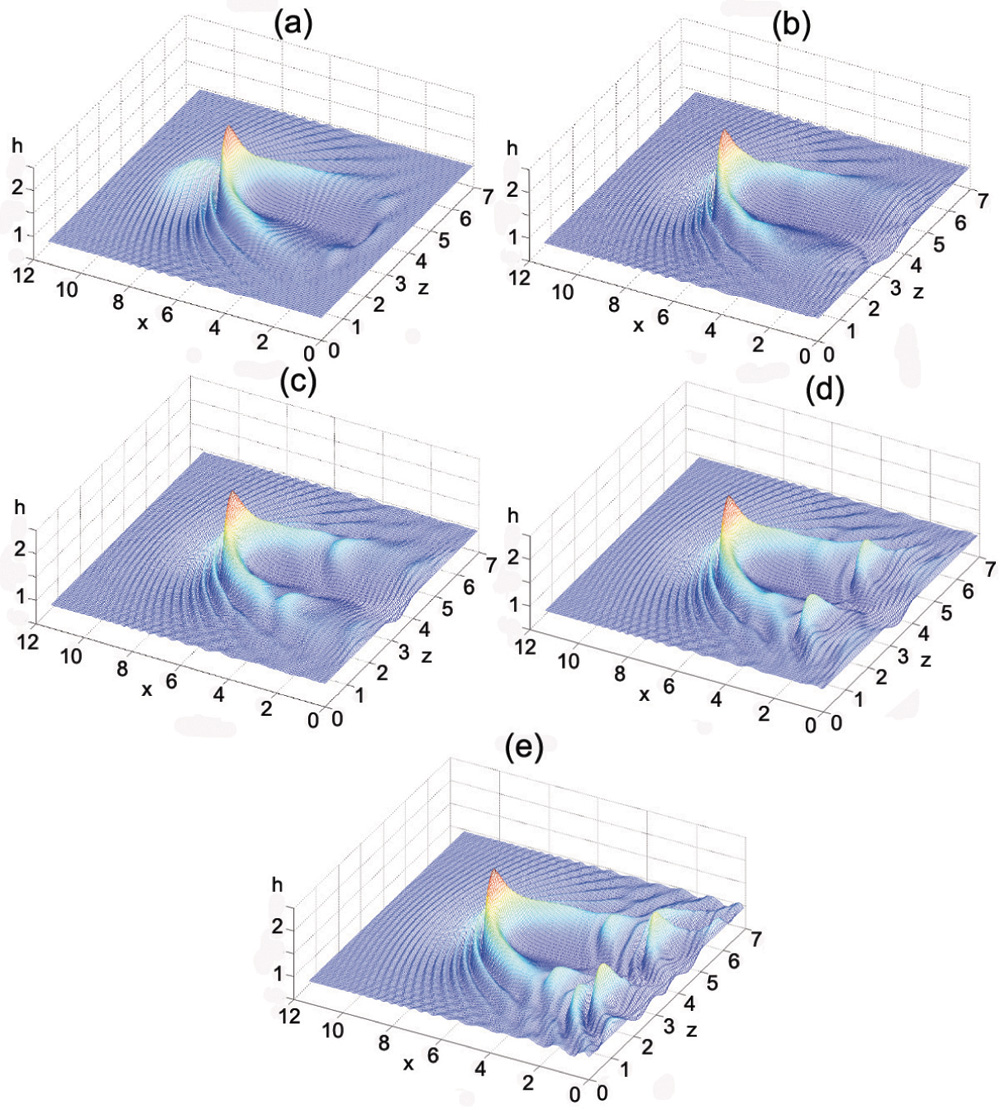}
\caption{Coalescence of a $\Lambda$-wave with a localized signal
of smaller amplitude and mass for $\delta=0.15$, (a) $t=0$ -
3D-soliton with a a localized signal in front of it; (b) $t=25$ -
the mass is absorbed by the 3D-soliton which becomes an
accelerated (excited) 3D-soliton; it is ready to swallow up an
equilibrium soliton; (c) $t=33$ - the extra-mass drains to the back
of  the soliton along its ``moustaches''; (d) $t=41$ - two small
$\Lambda$-waves created from the extra mass; (e) $t=45$ -
``checkerboard'' pattern starts to form.}\label{fig17}
\end{figure}


\newpage
\baselineskip18pt
\begin{thebibliography}{99}

\bibitem{D1}
Demekhin, E.A., Kalaidin, E.N., Kalliadasis, S. and Vlaskin, S.
Yu. Three-dimensional localized coherent structures of surface
turbulence. Scenarios of 2D-3D transition. {\em Phys. Fluids} {\bf
19}, 114103 (2007).

\bibitem{D2}
Demekhin, E.A., Kalaidin, E.N., Kalliadasis, S. and Vlaskin, S.
Yu. Three-dimensional localized coherent structures of surface
turbulence. $\Lambda$ solitons. {\em Phys. Fluids} {\bf 19},
114104 (2007).

\bibitem{P1}
Park, C.D. and Nosoko, T. Three-dimensional wave dynamics on a
falling film and associated mass transfer. {\em  AIChE J.} {\bf
49}(11), 2715 (2003).

\bibitem{A7}
Alekseenko, S.V., Antipin, V.A., Guzanov, V.V., Markovich, D.M.
and Kharlamov, S.M.  Stationary solitary three-dimensional waves
on a vertically flowing fluid film. {\em Doklady Phys.}, {\bf
50}(11), 598 (2005).

\bibitem{A8}
Alekseenko, S.V., Antipin, V.A., Guzanov, V.V., Kharlamov, S.M.
and Markovich, D.M.  Three-dimensional solitary waves on falling
liquid film at low Reynolds numbers. {\em Phys. Fluids}, {\bf 17},
121704 (2005).

\bibitem{LSG}
Liu, J., Schneider, J.B. and Gollub, J.P. Three-dimensional
instabilities on film flows. {\em Phys.\ Fluid}, {\bf 7}, 55
(1995).

\bibitem{L4}
Liu, J., Paul, D. and Gollub, J.P.  Measurements of the primary
instabilitites of film flows, {\em J. Fluid Mech.} {\bf 250}, 69
(1993).

\bibitem{V1}
Vlachogiannis, M. and  Bontozoglou, V. Observation of solitary
wave dynamics of film flows.  {\em J. Fluid Mech.} {\bf 435}, 191
(2001).

\bibitem{A5}
Alekseenko, S.V., Nakoryakov, V.E. and Pokusaev, B.G. Wave
formation on a vertical falling film. {\em AIChE J.} {\bf 31},
1446 (1985).

\bibitem{L2}
Liu, J. and Gollub, J.P. Solitary wave dynamics on film flows.
{\em Phys.\ Fluid}, {\bf 6}, 1702 (1994).

\bibitem{TSV}
Tihon, J.,  Serifi, K.,   Argyriadi, K. and Bontozoglou, V. Solitary
waves on inclined films: their characteristics and the effects on
wall shear stress. {\em Experiments in Fluids} {\bf 41}, 79
(2006).

\bibitem{C10}
Chu, K.J. and Dukler, A.E. Statistical characteristics of thin,
wavy films. Part 2. {\em  AIChE J.} {\bf 20}(4), 695 (1974).

\bibitem{A4}
Alekseenko, S.V., Nakoryakov, V.E. and Pokusaev, B.G. Wave flow of
liquid films.  {\em Begel House, Inc.}, 314 (1994).

\bibitem{A2}
Adomeit, F. and Renz U. Hydrodynamics of three-dimensional waves
in laminar falling films. {\em Int. J. Multiphase Flow} {\bf 26},
1183 (2000).

\bibitem{S6}
Selin, A.C. Experimental investigation of 3D wave structures in a
liquid film and mathematical modelling of its surface
instabilities.  PhD thesis, Perm (2009).

\bibitem{C7}
Chang, H.-C. and Demekhin, E.A. Complex wave dynamics on thin
films. {\em Elsevier}, 402 (2002).

\bibitem{D10}
Demekhin, E.A., Kalaidin, E.N., Shapar, S.M. and Shelistov, V.S.
About theory of multi-hump solitons in active-dissipative media.
{\em Izv. RAN, Mekh. Zhidk. i Gaza.} {\bf 2}, 186 (2009).

\bibitem{C5}
Chang, H.-C., Demekhin, E.A. and Kalaidin, E.N.  Interaction
dynamics of solitary waves on a falling film. {\em J. Fluid Mech.}
{\bf 294}, 123 (1995).

\bibitem{A9}
Aubry, N., Holmes, P., Lamley, J.L. and Stone, E.  The dynamics of
coherent structures in the wall region of a turbulent boundary
layer. {\em J. Fluid Mech.} {\bf 192}, 115 (1988).


\bibitem{D3}
Demekhin, E.A., Tokarev, G.Yu and Shkadov, V.Ya. On the existence
of critical Reynolds number for the falling by gravity liquid
film. {\em Teor. Osn. Khim. Tekhnol.} {\bf 21}(4), 555 (1987).

\bibitem{K3}
Kapitsa, P.L. Wave flow of thin viscous fluid layers, {\em Zh.
Eksp. Teor. Fiz.} {\bf 18}, 1 (1948).

\bibitem{K4}
Kapitsa, P.L. and Kapitsa, S.P. Wave flow of thin liquid layers.
{\em Zh. Eks. Teor. Fiz.} {\bf 19}, 105 (1949).

\bibitem{S4}
Shkadov, V.Ya. Wave modes in the gravity flow of a thin layer of a
viscous liquid. {\em Izv. Akad. Nauk SSSR, Mekh. Zhidk. i Gaza.}
{\bf 3}, 43 (1967).

\bibitem{S5}
Shkadov, V.Ya. Theory of wave flows of a thin layer of a viscous
liquid. {\em Izv. Akad. Nauk SSSR, Mekh. Zhidk. i Gaza.} {\bf 2},
25 (1968).

\bibitem{D5}
Demekhin, E.A. and Shkadov, V.Ya. Two-dimensional wave regimes of
a thin liquid film. {\em Izv. Akad. Nauk. SSSR, Mekh. Zhidk. i
Gaza.} {\bf 3}, 63 (1985).

\bibitem{D6}
Demekhin, E.A. and Shkadov, V.Ya. Theory of solitons in systems
with dissipation. {\em Izv. Akad. Nauk. SSSR, Mekh. Zhidk. i
Gaza.} {\bf 3}, 91 (1986).


\bibitem{C8}
Chang, H.-C., Demekhin, E.A. and Saprikin, S.S. Noise-driven wave
transitions on a vertically falling film. {\em J. Fluid Mech.} {\bf
452}, 255 (2002).






\end{thebibliography}
\end{document}